# A PROPOSED "OSI BASED" NETWORK TROUBLES IDENTIFICATION MODEL

Murat Kayri[1] and İsmail Kayri[2]

[1]Department of Computer and Instructional Technology, Yuzuncu Yil University, Van, Turkey
mkayri@yyu.edu.tr

[2] Department of Electric Education, Batman University, Batman, Turkey
ikayri@hotmail.com

*ABSTRACT*

*The OSI model, developed by ISO in 1984, attempts to summarize complicated network cases on layers. Moreover, network troubles are expressed by taking the model into account. However, there has been no standardization for network troubles up to now. Network troubles have only been expressed by the name of the related layer. In this paper, it is pointed out that possible troubles on the related layer vary and possible troubles on each layer are categorized for functional network administration and they are standardized in an eligible way. The proposed model for network trouble shooting was developed considering the OSI model.*

## 1. INTRODUCTION

As it is known, ISO (International Organization for Standardization) introduced the OSI (Open System Interconnection) standard in 1984. The system summarizing sophisticated network phenomena and cases on seven layers is called the OSI model. The OSI model deals with cases on the seven layers modularly revealed in a different fashion. The OSI model was basically developed to simplify network complexity, facilitate network training and introduce easy network troubleshooting [1]. The layers of the OSI model are shown in Figure 1:

| Application Layer |
| Presentation Layer |
| Session Layer |
| Transport Layer |
| Network Layer |
| Data Link Layer |
| Physical Layer |

Figure 1. OSI model

As it is clear in Figure 1, the OSI model consists of seven layers. Each layer concerns network cases at identification phase. In other words, each layer has different features and network cases or phenomena are expressed on these layers. Each layer provides data for the next layer. Any software and hardware components are tackled on the related layer [2, 3, 4].

### Application Layer (AL)

AL provides an interface between computer application and network. AL is the layer where network software (spreadsheet, word processor, FTP, TFTP, DNS, http and etc.) is generally defined.

### Presentation Layer (PRL)

This layer attempts to identify the type of data to be transferred to network settings. On this layer, whether the package to be transferred is a text, picture, video or sound file is identified. PRL regulates the nature and type of data to be transferred, so data format is defined.

### Session Layer (SL)

In SL, two computer applications are connected, used and ended. When a computer is in communication with more than one computer, SL provides communication with the right computer. In this process, data to be transferred to SL is separated by different sessions.

### Transport Layer (TL)

TL divides data from top layers into pieces in the size of a network package. TCP, UDP protocols start on this layer. The protocols carry out tasks like error control. TL serves as a means of top layer transfer and also increases quality of network service (QoS – Quality of Service). As it is known, data on this layer is divided into pieces called segments and gradually transferred to the lower layer. TL provides end-to-end data transfer. Error control is done and due data transfer is checked.

### Network Layer (NL)

NL is one of the most active layers in the OSI model. This is the layer where telecommunication services are active and router activities are defined. On this layer, data in segments are divided into packages and gradually transferred to the lower layer [6]. In NL, the most economical data transfer between two stations is controlled. Owing to this layer, data is directed through routers. Messages are addressed at network phase and also reasonable addresses are changed into physical addresses. At this phase, procedures such as network traffic and directions are completed. IP protocol starts on this layer.

### Data Link Layer (DLL)

On this layer, access methods such as Ethernet or Token Ring start. These access methods process and transfer data according to their own protocols [5]. On DLL, data is transmitted to Physical Layer from network layer. At this phase, data is divided into certain pieces which are called frames. Frames are packages which provide controlled data transfer. Most of DLL is performed within network card. The OSI model reports switching and bridging performed on DLL. DLL functions to identify other computers in network detect which user occupies the cable and control data from physical layer against errors.

### Physical Layer (PL)

PL defines the type of data on the line. Data is transferred as bits. This layer defines how to change 1s and 0s into electrical signals, light signals or radio signals and how to transfer them. On data sending side, physical layer changes 1s and 0s into electrical signals and places them on the line and on data receiving side, physical layer changes signals from the line back into 1s and 0s. In PL, there are active devices such as hubs and modems and passive components such as cable and 0-1(bits) [7].

As it is mentioned above, one of the most important functions of the OSI model is modular network troubleshooting. In the OSI model based on divide–split–swallow philosophy, network troubles are easily identified. In this way, troubleshooting is quick. Since the OSI model consists of seven layers, network troubles are dealt with under seven groups. However, when the fact that each layer includes a lot of details is taken into account, right trouble identification and troubleshooting is occasionally time-consuming. The main aim of this research is to identify possible network troubles by categorizing each layer in itself. In other words, each layer will be divided into a few categories according to troubles. In this way, troubles are realistically identified and "network troubles identification model" developed for the research could be used as a standard in network administration.

## 2. PROPOSED NETWORK TROUBLES IDENTIFICATION MODELS

### 2.1. Trouble Classification on Application Layer

As it is known, AL is a layer which includes all network software but it is irrelevant to hardware. Here, it is possible to divide software into two categories: i) **Off-line software** (word processor, spreadsheet and etc.), ii) **On-line software** (TFP, FTP, http, DNS and etc.). In this case, troubles on AL are divided into two: **"Off-line Troubles"** and **"On-line Troubles"**. In network administration, off-line troubles are problems inside the host, and on-line troubles are software problems which affect sharing between computers.

### 2.2. Trouble Classification on Presentation Layer

In network settings, there might be share troubles caused by file format. Basically, network data has three formats: text, audio, video–graphic. In this respect, problems might be categorized as: **"Text Troubles"**, **"Audio Troubles"** and **"Video-Graphic Troubles"**. Therefore, text troubles are format malfunctions of text only files in Word, Excel and etc. Similarly, audio troubles are possible problems in formats such as "wav" and "avi". Video-graphic troubles refer to malfunctioning file formats such as "mpeg", "jpeg", "tiff" and "bmp". Briefly, troubles on PRL are divided into three categories.

### 2.3. Trouble Classification on Session Layer

SL, which functions to start communication between two computers [8] and end communication when data share is over, sometimes serves for right communication when there is multi-connection. Troubles in peer-to-peer model with two computers only are grouped under one category and troubles in multi-connection are grouped under another . In this way, troubles on SL are divided into two: These are **"Peer-to-Peer Troubles"** and **"Multi-Connection Troubles"**.

### 2.4. Trouble Classification on Transport Layer

On TL, where data to be transferred is first divided, the number of moves needed for segment transfer to the next terminal. This could be performed as one-way synchronous, two-way synchronous or three-way synchronous. Dividing is possible with UDP (User-Datagram Protocol) or TCP (Transmission Control Protocol). UDP is unreliable but fast, whereas TCP is reliable but slow. In this context, troubles on TL could be divided into two: "UDP Based Troubles" and "TCP Based Troubles". Also, there might be troubles when buffers occupy segment transfer settings. When this case is considered, network troubles on TL are divided into three categories: **"UDP Based Troubles", "TCP Based Troubles"** and **"Buffer Troubles"**. In this way, troubles on this layer are identified at a micro level for network administration and troubleshooting is faster.

## 2.5. Trouble Classification on Network Layer

It is the most active layer in Network and is known as a setting where telecommunication services and router devices are active. IP, IPX/SPX, known as routed protocol, is covered on Network Layer. Furthermore, RIP (Routing Information Protocol), IGRP (Interior Gateway Routing Protocol), EIGRP (Enhanced Interior Gateway Routing Protocol), OSPF (Open Shortest Path First), AGP (Autonomous Gateway Protocol) and BGP (Border Gateway Protocol), known as routing protocol, are identified on Network Layer [9, 10, 11]. Echo message to realize package exchange (inbound-outbound) or to test next terminal reach is covered on this layer. As it is clear, router, routed and routing functions are performed on this layer. Therefore, troubles on this layer are needed to be categorized effectively. Standard for router hardware troubles might be called "**Hardware Troubles on NL**". Non-reach troubles by echo messages such as ping to the next terminal, and traceroute might be categorized as **"Echo Message Troubles"**. Troubles caused by telecommunication systems in general might be categorized as "**Telecommunication Troubles on NL**". Troubles on NL caused by wrong structuring of routed protocols are standardized as "**Routed Configuration Troubles**" and troubles on NL caused by wrong structuring of routing protocols are standardized as "**Routing Configuration Troubles**". As it is clear, troubles on NL in network administration cannot be handled under one category and the fact that troubles on one layer may vary should be taken into account. Thus, trouble standardization by categorization is inevitable.

## 2.6. Trouble Classification on Data Link Layer

On this layer, there are active devices such as switch and bridge used in LAN administration, and protocols such as PPP (Point-to-Point Protocol), HDLC (High Data Link Control), and Frame Relay which are Wide Area Network's protocols. As a result, troubles on this layer might belong to both LAN and WAN. On DLL, there is NIC (Network Interface Card), an indispensible part for LAN for communication. As it is easy to recall, there are two sub-layers on DLL: MAC (Media Access Control) and LLC (Logic Link Control). MAC totally refers to framing on LAN based on NIC, while LLC expresses the size of connection between DLL and NL. Standardization of troubles on DLL will be realized according to the above mentioned information. Troubles on WAN might be standardized as **"WANs Protocols Troubles on DLL"**. Troubles on NIC, an indispensible part for communication in LAN administration, are called **"NIC Troubles"** and troubles in MAC address, which has virtually been changed recently, are called **"MAC Troubles"**. Therefore, standardization is ensured. Troubles on LLC, which provides transfer from DL to NL, might be standardized as **"LLC Troubles"**. In this sense, it is possible for a network administrator to identify troubles on DLL immediately.

## 2.7. Trouble Classification on Physical Layer

It is known that all data on communication networks are changed into electrical signals (0-1). Data to be transferred in network settings are changed into electrical signals on PL and this layer is the last phase where data is divided. In this context, on PL, there are active-passive components such as electrical signals, cables, modems and hubs. Troubles on this layer are generally electrical problems. Troubles on cables such as UPT, STP, Coaxiel and RS-232, V.35, especially used in LAN and WAN, are handled on PL. Cable based troubles on PL might be categorized as **"Cables Troubles"** and excessive voltage loadings might be called **"Overload Voltage Troubles"**. In addition, troubles in hub devices used on LAN might be standardized as **"Hub Troubles"** and troubles in modems used on WAN might be standardized as **"Modem Troubles".** Moreover, broadcast troubles in wireless settings should be handled on PL. **"Wireless Wave Signal Troubles"** standard may be eligible for such troubles. In this way, possible troubles on PL are standardized and troubleshooting in network administration could be faster.

In the light of this information, the schematic approach in Figure 2 for network trouble standardization is obtained.

| LAYERS | TROUBLES |
|---|---|
| Application ➡ | Off-line Troubles<br>On-line Troubles |
| Presentation ➡ | Text Troubles<br>Audio Troubles<br>Video-Graphic Troubles |
| Session ➡ | Peer-to-Peer Troubles<br>Multi-connection Troubles |
| Transport ➡ | UDP Based Troubles<br>TCP Based Troubles<br>Buffer Troubles |
| Network ➡ | Hardware Troubles on NL<br>Echo Message Troubles<br>Telecommunication Troubles on NL<br>Routed Configuration Troubles<br>Routing Configuration Troubles |
| Data Link ➡ | WANs Protocols Trouble on DLL<br>NIC Trouble<br>MAC Trouble<br>LLC Trouble |
| Physical ➡ | Cables Troubles<br>Overload Voltage Troubles<br>Hub Troubles<br>Modem Troubles<br>Wireless Signal Wave Troubles |

Figure 2. Network Troubles Based on OSI Model

## 3. CONCLUSION

In this research, network troubles were identified and standardized for network administration. In this way, network troubleshooting could be identified faster. In the traditional approach, network troubles are expressed by names of layers. However, as it was shown in this research, troubles might vary on each layer. Therefore, troubleshooting will be faster. Such a standardization approach was OSI based and the proposed model was structured in this way. The model is thought to contribute to make network troubleshooting faster and easier.

## REFERENCES

[1]     Stewart, K., Adams, A. and Reid, A. (2008). *Designing and Supporting Computer Networks, CCNA Discovery Learning Guide*, Cisco Press, USA.

[2]     Diane, T. (1999). *Designing Cisco Networks,* Cisco Press, USA.

[3]     Rudenko, I. (2000). *Cisco Routers,* Coriolis Press, USA.

[4]     Giles, R. (1999). *All-in-one CCIE Study Guide,* McGraw Hill Press, USA.

[5]     Odom, S., Hammond, D. (2000). *Switching,* Coriolis, USA.

[6]     Larson, R.E, Low, C. S. and Rodriguez, P. (2000). Routing, Coriolis Press, USA.

[7]     Amato, V. (1999). *Cisco Networking Academy Program: Engineer Journal and Workbook Volume II*, Cisco Press, USA.Mason,

[8]     Mizanian, K, Vasef, M. and Analoui, M. (2010) "Bandwidth modeling and estimation in peer to peer networks", *International Journal of Computer Networks & Communications (IJCNC),* Vol. 2, No. 3, pp 65-83.

[9]     Yuste, A.J., Trivino, A., Trujillo, F.D., Casilari, E. And Estrella, A.D. (2009) "Optimized gateway discovery in hybrid manets", *International Journal of Computer Networks & Communications (IJCNC),* Vol. 1, No. 3, pp 78-91.

[10]    Moy, J.T. (1998). *OSPF Anatomy of an Internet Routing Prtocol,* Addison-Wesley Press, USA.

*learning with social software.* Retrieved 10.01.2010, from

http://www.dream.sdu.dk/uploads/files/Anne%20Bartlett-Bragg.pdf

[11]    Black, U. (2000). *IP Routing Protocols RIP, OSPF, BGP, PNNI & Cisco Routing Protocols,* Prentice Hall Press, New Jersey.

**Dr. Murat Kayri**

Dr. Kayri is an assistant professor in Computer Science and Instructional Technology Department in Yuzuncu Yil University. Dr. Kayri interests in neural network, statistical modelling, and networking. He has lots of articles on statistical and artificial neural network.

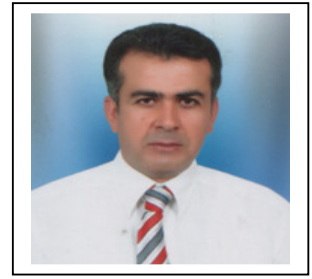

**PhD Scholar İsmail Kayri**

Kayri is both a lecturer in Department of Electric Education in Batman University and a PhD scholar in Electric-Electronic Department in Firat University. He studies on electric systems installation, programming language, database management system and other software tools especially related to electric-electronic science.

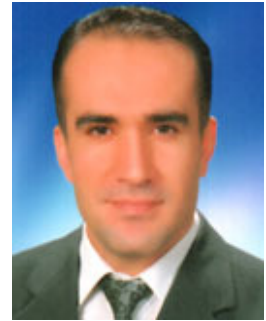